\documentclass[sigconf,pdfa]{acmart}
\newcommand{\eg}{\emph{e.g.}\xspace}
\newcommand{\ie}{\emph{i.e.}\xspace}
\newcommand{\OTHR}{Technical University of Applied Sciences Regensburg}
\newboolean{diff}
\setboolean{diff}{false}

\newcommand{\etal}{\emph{et al.}\xspace}

\usepackage{comment}
\usepackage{flushend}
\usepackage{csquotes}
\usepackage{placeins}
\usepackage{mathtools}
\usepackage{amsthm}
\usepackage{amsmath}
\usepackage{xcolor}
\usepackage{tabularx}
\usepackage{booktabs}
\usepackage{braket}
\usepackage{ifthen}
\usepackage[inline]{enumitem} 
\usepackage{xspace}
\usepackage[UKenglish]{babel}
\usepackage{censor}
\newboolean{anonymous}
\setboolean{anonymous}{false}
\ifbool{anonymous}{}{\renewcommand{\censor}[1]{#1}}
\ifbool{anonymous}{}{}
\ifbool{anonymous}{}{}

\newtheorem{example}{Example}[section]

\copyrightyear{2026}
\acmYear{2026}
\setcopyright{cc}
\setcctype{by}
\acmConference[Q-Data '26]{The third workshop on Quantum Computing and Quantum-Inspired Technology for Data-Intensive Systems and Applications}{May 31-June 05, 2026}{Bengaluru, India}
\acmBooktitle{The third workshop on Quantum Computing and Quantum-Inspired Technology for Data-Intensive Systems and Applications (Q-Data '26), May 31-June 05, 2026, Bengaluru, India}
\acmDOI{10.1145/3811628.3811835}
\acmISBN{979-8-4007-2703-0/2026/05}

\begin{document}

\title[A Toolbox to Understand the Physics of Data Management]{A Toolbox to Understand the Physics\\ of Quantum Data Management}
\author{\censor{Wolfgang Mauerer}}
\affiliation{
  \institution{\censor{\OTHR}}
  \institution{\censor{Siemens AG, Foundational Technology}}
  \city{\censor{Regensburg/Munich}}
  \country{\censor{Germany}}}

\author{\censor{Manuel Schönberger}}
\affiliation{
\institution{\censor{Cornell University}}
\city{\censor{Ithaca}}
\state{\censor{NY}
\country{\censor{USA}}
}
}

\renewcommand{\shortauthors}{\censor{Mauerer} and \censor{Schönberger}}

\begin{abstract}
The application of quantum computing to data management has attracted growing interest, yet remains constrained by a limited understanding of how the physical behaviour of quantum devices relates to the structure and difficulty of database problems. In particular, evaluating quantum annealing approaches for combinatorial optimisation, which is central to many data management tasks, poses significant challenges beyond the scope of conventional empirical and complexity-theoretic methods.

We present a computational toolbox for the systematic numerical analysis of quantum annealing processes derived from data management problem formulations. Adopting a physics-informed perspective, the toolbox enables the study of spectral and dynamical properties~--~such as energy gaps and eigenstate structure~--~that are inaccessible through direct hardware measurements, yet essential for understanding computational hardness and scaling behaviour.

Our approach further provides derived quantities and visualisation techniques that support the interpretation of optimisation dynamics, the identification of structural similarities to canonical physical models, and the construction of reduced effective descriptions. By bridging methodological gaps between quantum computing and database systems research, this work establishes a principled foundation for evaluating quantum approaches and guiding future co-design efforts.
\end{abstract}




\maketitle

\section{Introduction}
The use of quantum computing to accelerate or otherwise improve over classical baselines has seen a growing interest in the last years~\cite{yue:2023:qswa,eisert:2025,carbonelli:24:CaFeJu,bayerstadler:2021}. While limitations of quantum computers, especially their inability to process even moderately large amounts of classical data~\cite{Hoefler:2023,gogeissl:24:qdata}, clearly limit the possible use-cases in DBMS, many fundamental and long-standing problems in data management concern the solution of combinatorial optimisation problems that work on large conceptual search spaces without the need for large concrete data.  Discussions about scaling of and 
advantage from quantum combinatorial optimisation arise plentiful in the literature~\cite{Schoenberger2023ready,Trummer2016,Fankhauser2021,Fankhauser2023,Schoenberger2023,Nayak2023,Schonberger2023General,Trummer2025,Winker2023,Franz2024,Liu2025,Bittner2020,Bittner2020Hardware,Groppe2021,Fritsch2023,Gruenwald2023,Kesarwani2024,Gruenwald2023,Kesarwani2024,Calikyilmaz2023} (see also~\autoref{sec:related}). Predicting and understanding the
performance of quantum approaches is a generic problem~\cite{Lorenz:2025}: An empirical analysis is mostly impossible given the unavailability of large enough quantum hardware and the noisiness of current systems. A theoretical analysis requires an entirely different approach than for classical algorithms, and is computationally challenging and typically as hard as solving the subject problem in the first place.

\begin{figure*}[htbp]
    \includegraphics{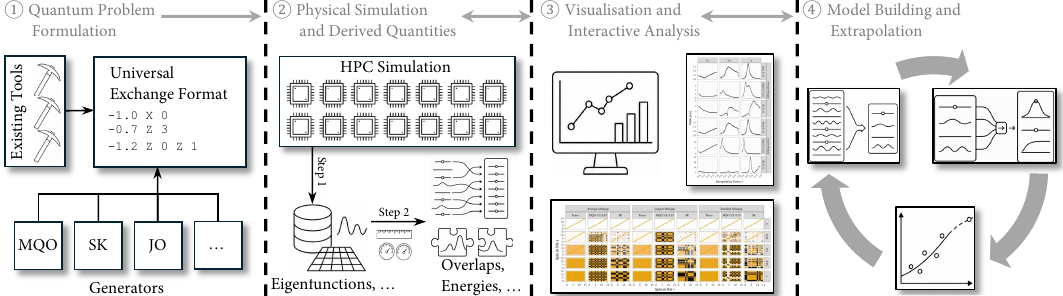}
    \caption{Overview of our Toolbox: Based on existing quantum formulations for optimisation problems that can be transcribed into a simple intermediate format, we use HPC systems to simulate the dynamics of a quantum annealing process. After post-processing the results (deliberately decoupled from the long-running and compute intensive simulation) to obtain quantities of interest, an accompanying set of visualisation scripts is used to understand the salient characteristics of the problem, and the implications on performance, achievable result quality, etc.; this can lead to (simplified) models, surrogates and others that allow for characterising properties of approaches, or predict characteristics.}\label{fig:overview}
\end{figure*}

In this paper, we present a tool for the classical numerical analysis of quantum annealing problems, and introduce, from a physics-informed perspective, a judiciously chosen set of 
quantities and visualisations that can be 
used to further the understanding of links
between the complexity-theoretic properties of such problems, and the physical properties of quantum annealers that are employed to solve them.
A general overview about our tool, available as fully reproducible~\cite{Mauerer2022} open source 
software on \censor{\url{https://github.com/lfd/anneal}}, is provided in~\autoref{fig:overview}.

While the theory behind many aspects of DBMS is extremely well developed, especially practical systems rely on a plethora of stochastic algorithms, heuristics or machine learning approaches. Any suggested quantum approach needs to be evaluated against such baselines, which is not possible empirically owing to the limitations of current prototypical machines. As the analysis of computational complexity (and other properties) of quantum annealing strongly differs from what is established in computer science, but is usually also impossible via a simple input-output relationship between problem and characteristics, new means of judging and predicting potential or expected performance of quantum approaches to DBMS problem are required.

Our goal is to establish a computational pipeline that allows researchers to address, among others, the following:
\begin{enumerate*}[label=(\alph*),itemjoin={{; }}, itemjoin*={{; and }}]
\item understand dynamical patterns and computational hardness of quantum approaches used for data management (and, more general, combinatorial optimisation) problems
\item compute quantitative and theoretically grounded estimates on scaling, success probability, etc. for such formulations, and enable performance predictions outside the empirically accessible problem range
\item reveal similarities of problem formulations in data management with seminal computational problems, also from the physics literature
\item construct surrogate simplified models based on
  effective dynamics that can be used to construct quantum-inspired approaches, or gain a deeper understanding of the properties of the original problem
\item avoid interpretation traps and pitfalls based on an incomplete or superficial 
  understanding of the underlying physical computational processes
\item identify possibilities for co-design approaches that allow for designing quantum accelerators with a focus on data management challenges
\end{enumerate*}

Based on our experience, we observe a certain amount of disconnect between established practices for theoretically and empirically assessing the guarantees and potential advantages of quantum approaches, and the expectations commonly applied in the evaluation of database systems research. In particular, methods that are standard and well-justified within the quantum computing literature do not always align with conventional benchmarking and evaluation paradigms in the DBMS community.
Work that follows established evaluation methodologies that may, for quantum approaches, not always be meaningful or appropriate, may be more readily accepted, whereas approaches that employ technically sound yet less familiar evaluation techniques can face additional scrutiny. One of the goals of this paper is therefore to raise awareness of this mismatch and to encourage a more nuanced discussion of evaluation criteria at the intersection of quantum computing and data management.

To judge hardness and scaling behaviour
in quantum optimisation requires considerably more knowledge
than can be obtained from simple empirical measurements
of success probabilities and quality of approximation measurements. For instance,
order parameters and their relation to occurring phase transitions are seen as crucial in the physics-centric literature, but rarely if ever employed for DBMS. Yet, the order of the underlying phase transition is a property of the spectrum and ground-state structure of 
\(H(s)\), not of one particular finite-rate run~\cite{Pelissetto2024}, which make more advanced and general empirical analysis techniques necessary. Our tool provides the necessary ingredients to conduct such analyses.

Even if the general overarching problem~--~using (future) quantum computers
to solve combinatorial optimisation problems~--~is not uniquely related
to problems in data management, we argue
that a co-design approach~\cite{Schoenberger2023ready,Li2021,Safi2025}, as well as careful tailoring of different
stages of the computational and conceptual pipeline from problem formulation
to systems integration is necessary to not only optimise benefits, but possibly also required to reach any form
of practical quantum advantage at all: The potentials of optimising
at the level of problem representation and transformation was, incidentally,
one of the early considerations in quantum data management~\cite{Trummer2016,Schoenberger2023},
but is increasingly addressed via automatic tools~\cite{Schmidbauer2025,Schmidbauer2026}, with
often considerable differences between algebraically identical formulations.
We expect that a better understanding of the underlying computational processes can provide
crucial input to such efforts.

While a considerable body of work addresses the
analysis of the annealing process as motivated by device physics, relating
problems in data management to this point of view is highly non-trivial and not yet sufficiently established. It is useful to be able to control the difficulty of constructed instances, and construct instances for arbitrary desired sizes. This is usually not the case for DBMS problems that are not continuous in the size of the generated Hamiltonians, and
display no immediate connection to Ising models.

\section{Concepts and Foundations}
As we discuss an interdisciplinary subject that goes
beyond the physical depth typically covered in 
the literature on quantum approaches for data management problems, 
we need to establish some foundations and consistent
terminology. While it is not possible to provide a self-contained introduction into
the physical and mathematical aspects of quantum mechanics, we nonetheless
summarise the most essential aspects that are required to understand our
approach. We assume familiarity with standard
concepts of quantum computing as increasingly adopted by the DBMS community.

\subsection{Quantum Annealing and Combinatorial Optimisation}

The Hamiltonian  \(\hat{H}\) is a Hermitian
operator that represents the total energy of a given system of interest. This can be any microscopic artefact like an atom, a molecule, or a constructed entity like a spin glass
whose properties are well suited to map combinatorial optimisation problems onto. The latter is the basis for
methods like quantum annealing that have received growing
interest in the data management community. While deriving
a Hamiltonian for a concrete problem is typically left to physics, it
is important to realise that it plays two key roles:
Firstly, the eigenvalues \(E_{i}\) of 
\(\hat{H}\) are the possible discrete energy levels\footnote{Hermiticity of \(\hat{H}\) guarantees a real eigen-spectrum, and thus real energy values.} that a given system can assume, each of which is accompanied by an eigenvector \(\ket{E_{i}}\), with \(\hat{H}\ket{E_{i}}=E_{i}\ket{E_{i}}\). The eigenvectors comprise an orthogonal set. Secondly,
the time evolution of a quantum system is governed by the time-dependent Schrödinger equation
\begin{equation}
i \hbar \frac{d}{dt} \lvert \psi(s(t)) \rangle = H(s(t)) \lvert \psi(t) \rangle,
\end{equation}
that contains the itself time-varying Hamiltonian \(\hat{H}(s(t))\) (we use a normalised time $s(t)$ defined by a monotonically increasing function \(s\) that runs from $0$ to $1$, and defines the annealing schedule) as a central ingredient. Knowledge of these two properties is essential to understand the behaviour of a system governed by a Hamiltonian, and therefore also
\emph{key to understanding the properties and behaviour
of computational approaches based on Hamiltonian dynamics}, 
particularly quantum annealing, but also related efforts like
QAOA. Obtaining these quantities to understand the effective computational process is our core goal.

Consider that at each value of $s$, the current state of the quantum annealer can be 
expanded in the \emph{instantaneous eigenbasis}
\begin{equation}
\ket{\psi(s(t))} = \sum_i c_i(t) \ket{E_i(s(t))}.
\end{equation}
Inserting this expansion into the Schrödinger equation  (essentially following textbook quantum mechanics, \eg~Ref.~\cite{Merzbacher1997}) yields coupled differential equations for the amplitudes $c_i(t)$, where transitions between eigenstates are governed by matrix elements of $\frac{d}{ds}H(s)$ and inversely by energy differences $E_{j}(s) - E_{i}(s)$. Note that energies and eigenstates change when the Hamiltonian 
changes, and re-computing the quantities is necessary.

This is particularly relevant for the class of problems arising 
in \emph{quantum annealing}, where the goal is to prepare the ground state (with the smallest possible energy \(E_{0}\) of a problem-specific Hamiltonian $\hat{H}_{P}$. 
This is, depending on the characteristics of \(\hat{H}_{P}\),
known to be a very hard computational problem~\cite{Farhi2000}.
Nonetheless, it can be solved by initialising the system in the ground state of a \emph{simple} \emph{driver Hamiltonian} $H_I$ 
with an easily prepared ground state, and then evolving the system according to an interpolation (a time-varying Hamiltonian)
\begin{equation}
\hat{H}(s) = A(s)\hat{H}_{I} + B(s)\hat{H}_{P} , \qquad s \in [0,1],\label{eq:anneal-hamiltonian}
\end{equation}
where  $H_P$ encodes the target optimisation problem (in the standard annealing protocol, \(\hat{H}_{I}=-\Gamma\sum_{i=1}^{n}\hat{\sigma}_{x}^{i}\) is used, with a known minimum-energy state 
\(\ket{+}^{\otimes n}\)). Quantum methods are often equated with quadratic unconstrained binary optimisation (QUBO) problems. More general choices are beyond our scope, but the link between QUBOs and \(\hat{H}_{P}\) is straightforward: A QUBO is a classical cost function
\(C(x) = \sum_i a_i x_i + \sum_{i<j} b_{ij} x_i x_j\) with \(x_i \in \{0,1\}\).
To embed this into a quantum annealer, binary variables are mapped to spins via
\(x_i = \tfrac{1}{2}(1 - \hat{\sigma}_{i}^{z})\), which converts the cost function into
a so-called Ising Hamiltonian \(\hat{H}_{P} = \sum_i h_i \hat{\sigma}_{i}^{z} + \sum_{i<j} J_{ij} \hat{\sigma}_{i}^{z} \hat{\sigma}_{j}^{z}\) (we equate spins and binary variables in the following).

This Hamiltonian is diagonal in the computational basis, and each bit-string encoded in a quantum state corresponds to a classical configuration with energy equal (up to a constant shift) to the original QUBO cost. Minimising the QUBO is equivalent to finding the ground state of \(\hat{H}_{P}\).

Note that the interpolation between initial and 
final Hamiltonian in~\autoref{eq:anneal-hamiltonian} is guided by the annealing parameter \(s\) that leads the system from $\hat{H}(0)=\hat{H}_{P}$ to $\hat{H}(1)=\hat{H}_{P}$ over an annealing time $T$. 
The schedule \(A(s), B(s)\)  determines how \(\hat{H}_{I}\) changes into \(\hat{H}_{P}\) over time. At $s = 0$, the system is initialised in the ground state $\ket{E_{0}(0)}$ of $\hat{H}_I$. The objective of quantum annealing is to evolve the system towards $s = 1$ such that the final state approximates the ground state $\ket{E_{0}(s=1)}$ of $\hat{H}_{P
}$. As
\(\hat{H}_{P}\) is designed such that a minimum-energy eigenstate
encodes the solution to a computational problem, the annealing
process can therefore be used to find solutions for this problem. In the ideal adiabatic regime, where $s(t)$ varies sufficiently slowly, transitions from the lowest-energy state into higher states are suppressed, and the system follows a single instantaneous eigenstate. In particular, if the system is initialised in \(\ket{E_{0}(0)}\),
then
\begin{equation}
\ket{\psi(t)} \approx e^{i\phi(t)} \ket{E_{0}(t)},
\end{equation}
up to a complex phase factor $\phi(t)$ that is inconsequential 
in a measurement.

Appropriately chosen schedules can not only influence how quickly, but also with which success probability the process reaches a
desired final state~\cite{Albash2018}, or which final energy state can be expected with which probability~--~which in turn influences solution quality (for the rest of this paper, we restrict our considerations to a standard linear schedule \(A(s) = (1-s)\), \(B(s) = s\); the tool supports arbitrary schedules though, as their computational handling is not much different from a linear schedule).
In contrast to the
adiabatic regime, \emph{diabatic} evolution refers to computations that involve transitions between eigenstates; the system explores excited states due to insufficient runtime or small gaps. This leads to computational results observed as incorrect solutions, albeit some algorithms deliberately exploit diabatic evolution~\cite{Crosson2021, Salatino2025, Muthukrishnan2016, Kadowaki2022}.

For each fixed \(s\), the Hamiltonian admits the aforementioned spectral decomposition in terms of the time-dependent energy eigenstates\footnote{While the use of braket notation is common in computer-science centric applications of quantum computing, the quantity \(\ket{E_{i}}\bra{E_{i}}\) is 
not ubiquitously used. For a finite-dimensional view of quantum mechanics, which is usually fully sufficient in computer science, \(\ket{E_{i}}\) corresponds to a complex column vector of dimension \(2^{n}\), whereas \(\bra{E_{i}}\) is the conjugate transpose, that is, a row vector. The outer product \(\ket{E_{i}}\bra{E_{i}}\) of a vector and its dual vector therefore corresponds to a \(\mathbb{C}^{2^{n}\times 2^{n}}\)  matrix that acts as an operator. As it is Hermitian by definition, it can be used as a contribution to a Hamiltonian.}
\begin{equation}
\hat{H}(s) = \sum_{i} E_{i}(s) \ket{E_{i}(s)}\bra{E_{i}(s)},
\end{equation}
where both eigenvalues and eigenstates depend smoothly on \(s\)
(under mild conditions). Given knowledge of \(\hat{H}_{I}\) and
\(\hat{H}_{P}\), numerical
methods can be used to obtain this decomposition, respectively the involved eigenstates and their eigenvalues. It is important to underline that performing an annealing
protocol on real quantum hardware does \emph{not} deliver this information, as usually
only probabilities of observing certain final states are recorded. Even
more involved measurement schemes do not deliver all information required
to obtain the spectral decomposition and the temporal dynamics. Additionally,
imperfect hardware~\cite{Albash2018,greiwe:23:imperfections,maschek:25:noise} as is available today will further perturb any empirical
observations, adding another degree of complications into the picture.

Consequently, numerical simulation of the process as performed by the tool
delivers \emph{more} information about the dynamics and the computational behaviour
as most empirical real-hardware studies.

While ideal adiabatic evolution tracks the ground state, 
practical implementations require considering a small number of 
low-energy eigenstates. For each $s$, define the truncated subspace
\begin{equation}
\mathcal{H}_{\mathrm{eff}}(s) = \mathrm{span}\{ \ket{E_0(s)}, \dots, \ket{E_k(s)} \}.
\end{equation}
Instead of requiring perfect adiabaticity, the state can be  approximated as
\begin{equation}
\ket{\psi(s(t))} \approx \sum_{i=0}^{k} c_{i}(s(t)) \ket{E_i(s(t))},
\end{equation}
with $k$ small. This viewpoint naturally arises when diagonalising $H(s)$ at discrete values of $s$ and propagating the state within the subspace spanned by a limited number of eigenstates, as is at the core of our numerical approach. The validity of this truncation depends on the structure of the spectrum along the interpolation path.

Another central quantity is the instantaneous spectral gap
\begin{equation}
\Delta(s) \coloneq \Delta_{10}(s) \coloneq E_1(s) - E_0(s),
\end{equation}
  More generally, \(\Delta_{k+1,k}(s) \coloneq E_{k+1}(s) - E_k(s)\) and \(\Delta_{k0}(s) \coloneq E_{k}(s)-E_{0}(s)\) denote gaps between higher energy levels that are of interest when considering larger subspaces and more involved dynamics, as they appear in real-world quantum annealers.
 If one of these gaps remains sufficiently large at some energy onwards, transitions to higher-energy states are suppressed, and the dynamics remain effectively confined to the effective Hilbert space $\mathcal{H}_{\text{eff}}(s)$ of the model. Likewise, \(\min_{s}(\Delta_{10}(s))^{-2}\) is known to be a first rough predictor for the required annealing runtime~\cite{Albash2018}. 
 
Let us again highlight that these spectral gap cannot be obtained from concrete measurements on a device without further ado, but are crucially relevant to determine and extrapolate scaling behaviour. This makes any attempt to perform empirical analyses on a concrete device necessarily incomplete.

The restriction to a low-energy subspace along the annealing path is justified when the following conditions hold:
\begin{itemize}
\item The system is initialised in the ground state of $H_I$.
\item The interpolation schedule $s(t)$ varies sufficiently slowly relative to inverse spectral gaps.
\item Energy gaps separating $\mathcal{H}_{\mathrm{eff}}(s)$ from higher-energy states remain non-negligible.
\item Matrix elements inducing transitions to higher-energy states are small.
\end{itemize}

Under these assumptions, high-energy components remain weakly populated. Moreover, any residual contributions acquire rapidly varying phases and have limited impact on observables, so that the effective dynamics are governed by a small number of instantaneous eigenstates.
Conversely, the approximation breaks down near avoided crossings with small gaps, under fast schedules, or in systems with dense spectra. In such cases, transitions out of the low-energy subspace become significant and must be explicitly accounted for.

From this perspective, quantum annealing can be interpreted as a controlled traversal of a sequence of low-dimensional eigenspaces of $H(s)$, where the computational task reduces to maintaining overlap with the evolving ground state (or a small set of low-energy states) of the interpolating Hamiltonian. As energy eigenvalues and eigenstates are of particular importance, careful consideration is required. Recall that at any fixed value of $s$, the Hamiltonian $H(s)$ has \emph{instantaneous eigenvalues}
\begin{equation}
E_0(s) \le E_1(s) \le E_2(s) \le \dots
\end{equation}
and corresponding \emph{eigenstates} $\ket{0(s)},\ket{1(s)},\dots$. A key subtlety is how the energy levels are labelled as $s$ varies. Two different conventions are commonly discussed: \emph{sorted energy branches} and \emph{tracked energy branches}. These correspond to two different ways of interpreting the spectral data, both supported by the tool.

In the sorted convention, eigenvalues are re-ordered at each $s$ 
so that $E_0(s)$ is always the lowest eigenvalue, $E_1(s)$ the 
second lowest, and so on. This ordering is local in $s$ and does 
not attempt to preserve the identity of the states across 
different values of $s$. The adiabatic theory of quantum 
annealing is formulated in terms of these sorted branches, since 
the evolution ideally follows the instantaneous ground state 
\(\ket{E_{0}(s)}\). Consequently, quantities such as the minimum 
gap \(\Delta_{\min} = \min_{s\in[0,1]} \Delta_{10}(s)\)
are defined using sorted energy levels.

Eigenstates can alternatively be tracked across different
values of $s$ by matching them according to overlap (\enquote{similarity}) between 
neighbouring eigenvectors (also illustrated in \autoref{fig:overview}). This produces \emph{continuous}
curves that represent the evolution of a specific physical state 
as the Hamiltonian changes. These tracked branches can cross 
each other and therefore do not necessarily remain ordered by 
energy. They are useful for interpreting how the character of 
the quantum state changes during the anneal and for identifying 
tunnelling between configurations.

Near an \emph{avoided crossing}, two tracked branches approach 
each other closely. In the 
sorted picture the labels of the levels exchange roles, whereas 
in the tracked picture the curves cross. This distinction is 
important: performance metrics such as the minimum gap or 
adiabatic condition must be computed using the sorted 
eigenvalues, while tracked branches are primarily used to 
visualise the physical structure of the transition.

\subsubsection{Scalability}
The problem we need to solve places considerable computational demands that cannot be provided on typical machines but for the smallest instances (and is also more complex than the widely used
simulation of quantum circuits), but requires the use of HPC methods. We 
rely on established and proven approaches:
Scalability is achieved through MPI parallelism that especially distributes huge state vectors across processes; 
OpenMP threading that allows for shared-memory parallelism within each process, and most importantly matrix-free evaluation
techniques minimises memory bandwidth requirements.
We observe good scalability of MPI-level parallelism with up to
about 150 cores; this is sufficient to handle one single problem formulation with 25 qubits in roughly one full day.
The required time for an analysis with eight
eigenpairs and 200 steps of resolution for more moderate system dimensions up to 20 qubits, as well as the storage requirements, are shown in~\autoref{tab:size} to familiarise researchers with the expected resource requirements.
Beyond around 150 cores, communication and parallelisation overheads eliminate further computational gains,
a resorting to parameter-level parallelisation is 
required. This can be conveniently established for sweeps
and parameter scans.

\begin{table}[htbp]
\caption{Storage requirement \(|E|\) for eigenpairs at different system dimensions (\# of qubits) \(N\), and for a complete system analysis \(|S|\) with eight eigenvectors computed at a resolution of 200 steps. Ensemble evaluations in the paper consider 100 instances at each dimension. Compute time \(t\) specifies the walltime for one instance of the SK problem in minutes.}\label{tab:size}
\begin{tabular}{lrrrrrrrrrrr}
\toprule
n  & 10 & 12 & 14 & 16 & 18 &  20\\
 \midrule
\(|E|\) [MiB] & 0.016 & 0.04 & 0.136 & 0.52  & 2.06  & 8.20\\
\(|S|\) [GiB] &  0.026 & 0.066 & 0.223 & 0.853 & 3.37 & 13.4\\
\(t\) (32 cores) & 0.08 & 0.15 & 0.62 & 2.01 & 12.3 & 74.7\\
\(t\) (64 cores) & 0.12 & 0.20 & 0.68 & 1.50 & 8.20 & 54.9\\
\(t\) (96 cores) & 0.24 & 0.35 & 4.26 & 1.75 & 7.10 & 28.0\\
\bottomrule
\end{tabular}
\end{table}

\subsubsection{Tool Use}
Hamiltonians respectively problems are specified as text files
with one term per line in format
\texttt{coeff op1 i [op2  j]}. For instance:

\begin{verbatim}
-1.0 X 0
-0.7 Z 3
-1.2 Z 0 Z 1
\end{verbatim}

Each term represents a Pauli operator acting on one or two qubits
(while it is possible to consider X and Y interactions, QUBO-based
formulations and standard Ising annealing problems will only
resort to using Z operators, and use X operators only implicitly for the initial Hamiltonian). The format is deliberately simple, and 
designed to easily interface with all available mechanisms for 
generating QUBO problems, regardless of previously utilised 
programming language or quantum framework. The most relevant
parameters of the tool are collected in~\autoref{tab:tool-params}.

\begin{table}[htbp]
\caption{Main parameters of the simulation tool. Standard parameters that set output paths etc., as well as parameters that control options like tolerance settings for the employed numerical techniques are not documented here, but are available through the source
code respectively the accompanying documentation.}\label{tab:tool-params}
\centering\begin{tabularx}{\linewidth}{lX}
\toprule
Parameter & Description \\
\midrule
\texttt{-N} & Number of qubits \\
\texttt{-nev} & Number of eigenpairs to compute \\
\texttt{-s\_start}, \texttt{-s\_end} & Range of interpolation $s$ \\
\texttt{-s\_steps} & Number of sampling points \\
\texttt{-HI\_file}, \texttt{-HP\_file} & Files for driver and problem Hamiltonian\\
\texttt{-auto\_generate\_hi} & Generate $H_I = -\Gamma \sum_{i} X_{i}$ \\
\texttt{-hi\_gamma} & Strength of driver Hamiltonian \\
\texttt{-track\_by\_overlap} & Enable eigenstate tracking \\
\texttt{-track\_observables} & Compute $\langle Z_i \rangle$\\
\texttt{-track\_zz\_correlations} & Compute $\langle Z_i Z_j \rangle$ \\
\texttt{-save\_eigenvectors} & Store eigenvectors \\
\bottomrule
\end{tabularx}
\end{table}

\begin{figure*}[htbp]
    \includegraphics{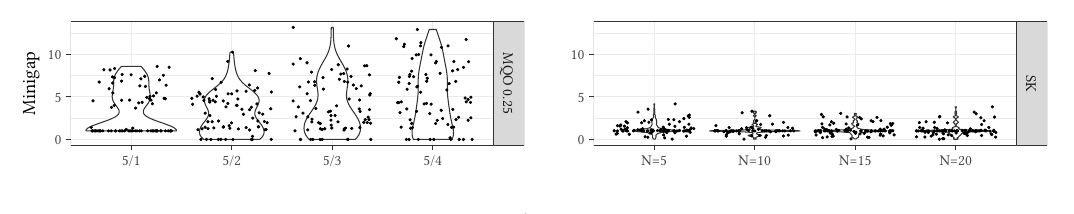}\vspace*{-3em}
    \caption{Distribution of minimum gap values for two subject problems: Multi-Query Optimisation from data management as a combinatorial optimisation problem of domain interest, and the Sherrington-Kirkpatrick spin-glass problem as canonical archetype for hard annealing problems.}\label{fig:minigap-dist}
\end{figure*}

\subsubsection{Derived Quantities and Post-Processing}
Primarily, the simulation produces eigenvalues $E_n(s)$ for $n=0,\dots,k-1$, and the corresponding eigenvectors. For each eigenstate \(\ket{E_{n}}\), the tool computes
the expected value of each spin \(\langle Z_i \rangle = \bra{E_{n}(s)} Z_i \ket{E_{n}(s)}\) (corresponding to the
local magnetisation, which is the expected value of a binary optimisation variable with values \(\{-1,1\}\)), and the correlation between any
pair of two spins \(\langle Z_i Z_j \rangle = \bra{E_{n}(s)} Z_i Z_j \ket{E_{n}(s)}\). These quantities provide insight into the structure of quantum states and their evolution on a path
towards the solution for a combinatorial optimisation problem,
as we discuss below.

Based on energy eigenvectors, a post-processing step (deliberately 
implemented as a second flexible step after the compute-intensive
diagonalisation) included in the tool can, among others,
compute energy gaps and matrix elements
\begin{align}
H_{mn}(s) &= \bra{E_{m}(s)}\hat{H}(s)\ket{E_{n}(s)}, \\
M_{mn}(s) &= \bra{E_{m}(s)}\partial/\partial_s \hat{H}(s)\ket{E_{n}(s)},
\end{align}
where \(\partial/\partial_{s} \hat{H}(s) = \hat{H}_{P}-
\hat{H}_{I}\) for the annealing Hamiltonian with linear 
schedule. These quantities
have various uses in analysing the performance of annealing
approaches. For instance, they are required when the
type of phase transitions that  for specific problems
in an annealing process are determined, and are~--~in particular \(M_{10}\)~--~central to estimating minimally required annealing times that ascertain the desired adiabaticity. Likewise, it is possible to construct  \emph{reduced models} like
\emph{two-level models} based on $(E_0, E_1, M_{10})$ (so-called Landau-Zener models that are attractive for their ultimate
simplicity, but usually insufficiently accurate for annealing dynamics),
or more general \emph{multi-level transport models} that
use a truncated subspace to describe the effective dynamics using a simplified surrogate description (see Ref.~\cite{Merzbacher1997,Albash2018,Roland2002,Amin2009b}).\footnote{In the instantaneous energy
eigenbasis, the evolution is known to obey:
\begin{equation}
i \frac{d}{dt} c_m(t) = E_m(s) c_m(t) - i \dot{s} \sum_n M_{mn}(s) c_n(t).
\end{equation}
This provides a low-dimensional approximation of the system dynamics; all required information to construct such models is available from the 
tool results.} From such models, it is then possible to compute, for instance, the ground-state probability $p_{\text{gs}}(T)$
or Time-to-solution (TTS)~\cite{Mehta2025}. Such metrics 
can connect spectral properties to algorithmic performance. 

Note, however, that the purpose of our tool and this paper is \emph{not} to construct and evaluate such models in detail~--~this is left for later work. Here, we set the stage for obtaining all required information for quantum problems relevant for data management. 

\subsection{Multi-Query Optimisation}\label{sec:mqo}
Multiple query optimisation (MQO) concerns the identification of a plan configuration that minimises the aggregate execution cost of a batch of queries evaluated jointly~\cite{Sellis1988}. In several formulations, the problem is NP-hard~\cite{Trummer2016}. A naïve optimiser may greedily select the individually cheapest plan for each query, thereby neglecting overlaps and potential cost-sharing opportunities across queries. Such locally optimal choices can be globally suboptimal. By contrast, MQO explicitly exploits common sub-expressions and inter-plan cost savings, seeking a configuration that minimises total cost while balancing individual plan expenses against shared savings.

Recent work by Schönberger~\etal~\cite{Schoenberger2025} demonstrates an approach to MQO based on a QUBO formulation, targeting quantum annealers and quantum-inspired hardware, with encouraging empirical results on the latter. As the problem can be scaled down to instances that are tractable with reasonable effort in our toolbox, and as the empirical results suggest that the problem is a good candidate for a neither too easy nor excessively hard problem in a QUBO-base formulation, we use it as reference case.

Formally, an MQO instance comprises: (i) a set of queries $Q$, (ii) a set of candidate execution plans $P$, and (iii) a set of cost-saving opportunities $S$. Each query $q \in Q$ is associated with a subset $P_q \subseteq P$ of mutually exclusive plans. Each plan $p_i \in P$ incurs a cost $c_i > 0$, which may be reduced through a saving $s_{p_i,p_j} \geq 0$ when combined with another plan $p_j$ from a different query. We denote instances as MQO $|Q|/|P|, d$, where $d$ captures the density of cost-sharing opportunities given by the fraction of cost savings featured by an MQO instance over all possible cost savings. For example, MQO $5/3, 0.25$ describes five queries, three plans per query, and a sharing density of 0.25.

\begin{example}
\label{example:mqo_problem}
 Consider an MQO scenario featuring four queries with plans $\mathit{P_{q_1}}=(p_1, p_2), \mathit{P_{q_2}}=(p_3, p_4), \mathit{P_{q_3}}=(p_5, p_6)$ and $\mathit{P_{q_4}}=(p_7, p_8)$. Further, let their execution costs be $c_1=9, c_2=10, c_3=9, c_4=10, c_5=11, c_6=9, c_7=14, c_8=9$, and the respective cost savings $\mathit{s_{p_1, p_3}}=1$, $\mathit{s_{p_1, p_4}}=1$, $\mathit{s_{p_2, p_3}=1}$, $\mathit{s_{p_2, p_4}=5}$, $\mathit{s_{p_2, p_7}=5}$, $\mathit{s_{p_4, p_5}}=5$, $\mathit{s_{p_5, p_7}=5}$, $\mathit{s_{p_5, p_8}}=1$, $\mathit{s_{p_6, p_7}=1}$, $\mathit{s_{p_6, p_8}}=1$.

 Greedily, a query optimiser may choose the locally cheapest plan for each query, resulting in the overall plan selection $\mathit{P_{gr}} = (p_1, p_3, p_6, p_8)$. Accounting for costs savings $\mathit{s_{p_1, p_3}}=1$ and $\mathit{s_{p_6, p_8}}=1$, we obtain total costs $C(\mathit{P_{gr}})=9+9+9+9-1-1=34$. However, by directly accounting for such cost saving opportunities during the search process, the optimiser instead arrives at the globally optimal plan configuration, given by $\mathit{P_{opt}} = (p_2, p_4, p_5, p_7)$, yielding total execution cost $C(\mathit{P_{opt}})=10+10+11+14-5-5-5-5=25$.
\end{example}

As our goal in this paper is
to show \emph{approaches} how to set data management problems in perspective with the performance of quantum annealing, we deliberately leave an actual analysis in more depth, as well as the consideration of additional DBMS problems,  to follow-up work for which this paper sets the stage.

\begin{figure*}[htbp]
    \includegraphics{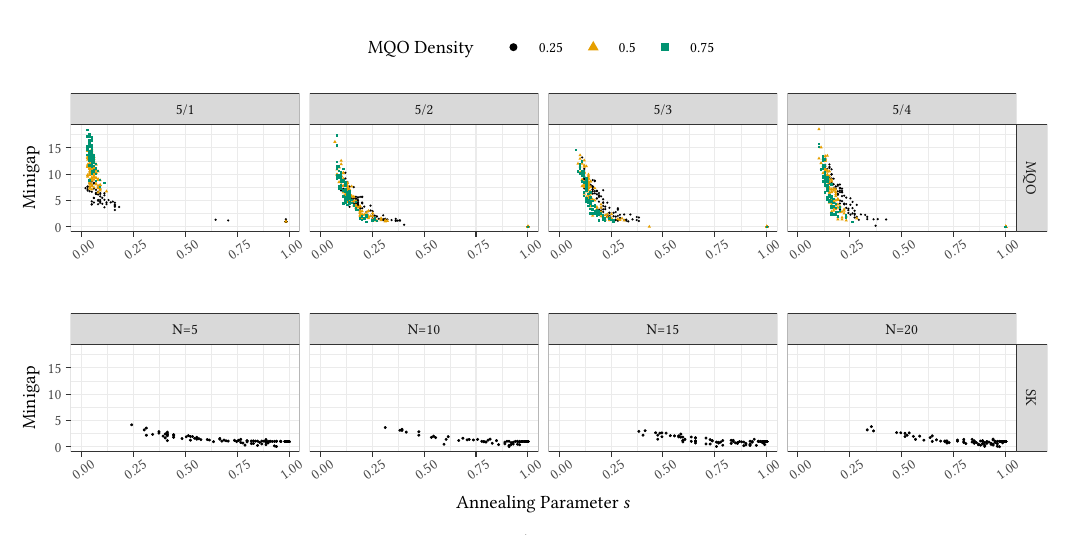}\vspace*{-2em}
    \caption{Minimum gap size versus temporal occurrence in the anneal process (at interpolation factor \(s)) \) for 100 random instances each of SK+RF and MQO. Labels N/M for MQO denotes N queries and M plans per query; qubit counts are vertically identical across panels.}\label{fig:minigap-scatter}
\end{figure*}

\section{Methods \& Approach}\label{sec:methods}
\subsection{Numerical Analysis}
The Hilbert space of an $n$-qubit system has dimension $2^n$, which grows exponentially with increasing problem dimension. Explicit matrix representations of $H(s)$ are therefore infeasible beyond very modest system sizes; even the largest supercomputers that are currently available can only simulate quantum systems with up to 50 qubits~\cite{DeRaedt2025}.
Instead, the computational problem is formulated as:

\begin{itemize}
\item Given $H_I$ and $H_P$ in operator form, compute the lowest $k$ eigenvalues and eigenvectors of $H(s)$ for a set of values $s_1, \dots, s_M$ (typically, we choose \(k \in [6, 8]\) to
balance accuracy with computational economy).
\item Extract derived quantities such as
instantaneous energy gaps  $\Delta_{1 0}(s) = E_1(s) - E_0(s)$,
expectation values $\langle Z_i \rangle$ and correlations $\langle Z_i Z_j \rangle$,
and matrix elements $\bra{E_{m}(s)}\hat{H}\ket{E_{n}(s)}$ and $\bra{E_{m}(s)}\partial_{s
}\hat{H}\ket{E_{n}(s)}$. All these are relevant to determine qualities and behaviour of the annealing
process for a given combinatorial optimisation problem.
\end{itemize}

This leads to a sequence of large-scale sparse eigenvalue problems. that must be solved as efficiently as possible. In particular, the Hamiltonian is never materialised explicitly as a dense matrix. Instead, it is represented as a \emph{matrix-free operator} implemented using PETSc's \texttt{MatShell} abstraction. The action of $\hat{H}(s)$ on a state vector is computed on-the-fly by applying Pauli operators to bit-string representations of basis states. This avoids the $\mathcal{O}(2^{2N})$ memory requirement of dense matrices and reduces storage to $\mathcal{O}(N)$ per term.
Our implementation is mainly based on the library \href{https://petsc.org}{PETSc}~\cite{petsc-efficient,petsc-user-ref} for distributed vectors, parallel communication, and matrix abstractions,
and \href{https://slepc.upv.es/}{SLEPc}~\cite{Hernandez:2005:SSF,Hernandez:2003:SSL,Hernandez:2007:PAE,Roman:2023:ISR} for iterative eigenvalue solvers. Eigenpairs are computed 
using the Krylov-Schur subspace method~\cite{Stewart2002}, which is well-suited for extracting a 
small number of extremal eigenvalues~\cite{Golub2013}. In our simulation examples, we 
computed the first eight eigenvalues (in the sense of smallest algebraic value), 
together with the associated eigenvectors. To ensure continuity of eigenstates 
across $s$, an \emph{overlap-based tracking} scheme is employed: We match 
eigenvectors obtained at discrete time $s_{k}$ to those at discrete time $s_{k-
1}$ by maximising overlaps.
This avoids artificial discontinuities from  crossings or solver reordering.

\subsection{Reference Problems}
Apart from the key subject problem of multi-query optimisation that we use to illustrate tool capabilities, our goal is
to relate hardness and complexity of quantum problem formulations in data management to existing results. Therefore, we have chosen a few reference problems with known properties that have been investigated in the literature for sometimes over decades.
All problems are formulated as instances of an Ising Hamiltonian \(\hat{H} = \sum_{i<j} J_{ij} s_i s_j \;+\; \sum_i h_i s_i\) with \(s_{i} \in \{\-1, +1\}\)
that corresponds directly to the cost function minimised by a quantum annealer at the end of the annealing schedule. Different problems are encoded by suitable choosing \(J_{ij}\) and \(h_{i}\); a direct relation between pseudo-Boolean functions~\cite{Schmidbauer2025,Schmidbauer2026} and Ising Hamiltonians is obvious.

\paragraph{Ferromagnetic Ising Model}
The ferromagnetic Ising model (note that different conventions for terminology are in use in the literature) serves as a tractable reference with well-understood scaling properties. We consider \(N\) spins
governed by the Hamiltonian 
\begin{equation}
H_{\text{FIM}} = - J \sum_{(i,j)\in E} s_i s_j -h\sum_{i}s_{i}.
\end{equation}
All couplings \(J>0\) are identical and favour alignment of spins. Local terms \(h\) are identical for every spin.

The ground states without local fields (\ie, \(h=0\)) are the two fully aligned configurations \(\mathbf{s} = (+1,\dots,+1)\) and \(\mathbf{s} = (-1,\dots,-1)\), with energy
\(E_{0} = -J |E|\). A small local bias \(h_{i}\) is used to break the symmetry between these 
two, which also eliminates some computational challenges. Excitations correspond to domain walls, and their energy cost
scales with the boundary size of flipped regions. For a complete graph, the energy gap between the ground state and first excited state scales as \(\Theta(N)\), yielding a simple energy landscape without local minima.

From a quantum annealing perspective, the problem
exhibits a second-order phase transition with a polynomially closing minimum gap, making this problem computationally easy. Consequently, it serves as a sanity check for correct scaling behaviour and for validating that the simulation reproduces known analytical results.

\paragraph{Sherrington-Kirkpatrick type models}
The Sherrington-\hspace*{0mm}Kirkpatrick (SK) model 
with random external field (SK+RF; we only refer to this scenario in the following)
is a paradigmatic problem. The Hamiltonian is defined as:
\begin{equation}
H_{\text{SK}} = - \sum_{i<j} J_{ij} s_i s_j  -\sum_{k}h_{k}s_{k},
\end{equation}
where the couplings \(J_{ij}\) and the local fields \(h_{k}\) are drawn from a Gaussian or uniform distribution.

Appropriate normalisation ensures an extensive energy (\ie, \(E = \Theta(N)\)). Unlike the ferromagnetic case, the interactions are frustrated: for a typical triple \((i,j,k)\), it is impossible to satisfy all pairwise interactions simultaneously. This leads to a highly rugged energy landscape. 

The SK model exhibits well-studied physical phenomena like a phase transition at a critical, separating a paramagnetic phase from a glassy phase characterised by replica symmetry breaking. From an optimisation viewpoint, this corresponds to a proliferation of local minima separated by extensive barriers.

In the quantum setting, the problem is believed
to exhibit exponentially small minimum spectral gaps in the worst case, leading to exponentially long annealing times. The combination of disorder, frustration, and complex phase structure results in an archetypical hard benchmark problem
for classical and quantum optimisation.

\paragraph{Hamming Weight Problem}
As a complementary control problem, we consider a Hamming-weight minimisation instance, which has a trivial structure and a constant spectral gap under quantum annealing. The Hamiltonian \(H_{\text{HW}} = \sum_{i=1}^{N} \frac{1 - s_{i}}{2}.\) is purely local;
the ground state is the all-\(+1\) configuration, with energy \(0\), and each spin flip increases the energy by exactly one unit. The spectrum is fully determined by the Hamming weight of the configuration:
\(E = \left|\{\, i \mid s_i = -1 \,\}\right|\). For quantum annealing,  the system decomposes into \(N\) independent single-qubit problems. The minimum spectral gap is constant and does not shrink with \(N\), implying that the annealing time required to reach the ground state with high probability is independent of problem size.
This problem provides a correctness and calibration benchmark; it is included
in the tool, but we do not discuss the expected trivial results in this paper.




\begin{figure}[htbp]
    \includegraphics{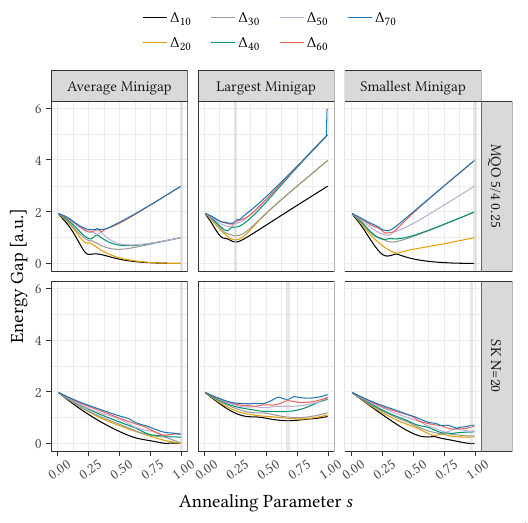}\vspace*{-0.5em}
    \caption{Energy curves obtained by point-wise sorting of \(s\)-local energy levels. These data are employed in computing quantities like the spectral gap \(\Delta_{10}(s)\) (vertical gray bar), and are used in runtime analyses. For the subjecty problems, a clearer separation between (non-degenerate) energy levels for MQO in contrast to SK is visible.}\label{fig:eb-sorted}
\end{figure}

\begin{figure}[htbp]
    \includegraphics{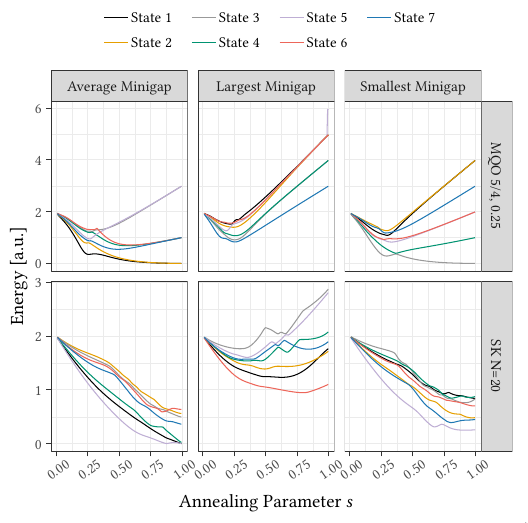}\vspace*{-0.5em}
    \caption{Energy curves for different states of the system obtained by tracking propagation from initial minimum energy overlap states. The extra effort beyond inferring
    point-wise sorted quantities in~\autoref{fig:eb-sorted} can provide additional insights into the physical properties of an annealing process.}\label{fig:eb-tracked}
\end{figure}

\section{Results and Discussion}\label{sec:results}
In discussing typical results, visualisations and interpretations obtainable from our tool, we
highlight salient differences between properties and behaviour of MOQ and SK as a known-hard reference problem, we focus on a general understanding of the meaning of the graphs, and do at this stage not proceed with a more fine-grained and detailed analysis of a specific problem
relevant in data management contexts.

\subsection{Instance Selection and Spectral Gaps}
\begin{figure}[htbp]
    \includegraphics{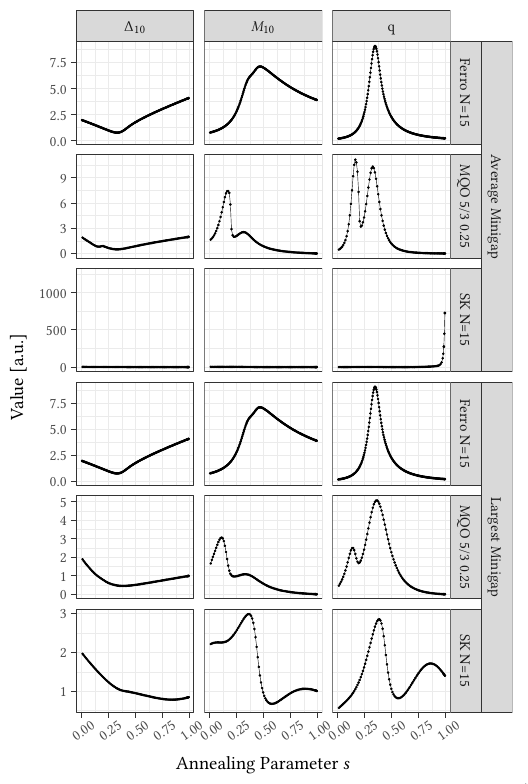}\vspace*{-0.5em}
    \caption{Temporal evolution of spectral minimum gap between 
    ground state and first excited state energy, transition 
    matrix element \(M_{1 0} = |\bra{E_{1}
    (s)}\partial_{s}\hat{H}\ket{E_{0}(s)}|\), and the so-called 
    adiabatic condition ratio 
    \(q=M_{10}/\Delta_{10}^{2}\). The quantities provide crucial insights into dynamical and hardness properties of the
    annealing process; as discussed in the text.}\label{fig:characteristic-dynamics}
\end{figure}

The properties of a given problem depend on the properties of
individual instances. In general, two points of view are possible:
Considering individual instances, or considering averaged ensemble properties. Representative and meaningful results require choose a representative ensemble, and to construct states with relevant properties. To ascertain the former, we use a randomly (but reproducibly) generated sample of 100 instances for each size of a given problem, and compute/visualise summaries, as  in~\autoref{fig:minigap-scatter}. Our choice of individual instances is based on the previously minimum spectral gap \(\min_{s}\Delta_{10}(s)\) of an instance, which is
a first-order proxy for minimal required annealing time  (inverse quadratic) and thus also instance hardness. From the ensemble of all instances, we choose the instance with the smallest and largest minimum gap, representing a hard and easy instance, as well one \enquote{typical} instance with average minimum gap in between the two boundary cases (see, \eg,~\autoref{fig:characteristic-dynamics}). While this does not entail a strict correspondence with complexity-theoretical hardness, it provides an unbiased selection of relevant instances.

\autoref{fig:minigap-dist} provide an initial overview
about these minimal spectra gaps across the subject instances; 
\autoref{fig:minigap-scatter} displays the same data in a more interesting 
way, as it compares the minimum spectral gaps and their locations along the 
annealing parameter \(s\) for SK and MQO. For the Sherrington–Kirkpatrick 
model (bottom row), increasing the system size leads to a clear trend toward 
smaller minimum gaps and a shift of the dominant avoided crossings toward 
later stages of the anneal, consistent with the emergence of many nearly 
degenerate classical configurations that induce small avoided level crossings 
(see also \autoref{fig:eb-sorted} and the following discussion). In contrast, 
the multi-query optimisation instances (top row) exhibit systematically larger 
gaps and earlier bottlenecks, indicating that the hardest spectral 
rearrangements occur while the initial driver Hamiltonian \(\hat{H}_{I}\) 
still plays a significant role. Importantly, annealing dynamics are governed not just by gap size but by the structure 
and location of avoided crossings. These are highly instance-dependent: SK 
realises a dense set of late-stage near-degeneracies, while MQO shows fewer 
and earlier crossings with larger gaps.

\subsection{State Evolution}
The evolution of sorted and tracked energy branches, as defined above, is show in~\autoref{fig:eb-sorted} and \autoref{fig:eb-tracked} for SK and MQO. The MQO instance exhibits a relatively smooth spectral evolution, with finite gaps throughout the annealing schedule and only mild avoided crossings, suggesting the absence of severe adiabatic bottlenecks at this system size. In contrast, the SK model displays multiple narrow avoided crossings and pronounced gap suppression near intermediate values of \(s\), consistent with its frustrated energy landscape and dense spectrum of competing states. While these observations align with expectations that SK-type problems become exponentially hard in the large-\(N\) limit (which is the reason why we have chosen them as a reference problem), the present finite-size results primarily indicate qualitative differences in spectral structure rather than definitive asymptotic scaling. The tracked eigenstate trajectories further reveal the locations of dynamical bottlenecks that are not fully captured by point-wise sorted gaps.

\begin{figure*}[htbp]
    \includegraphics{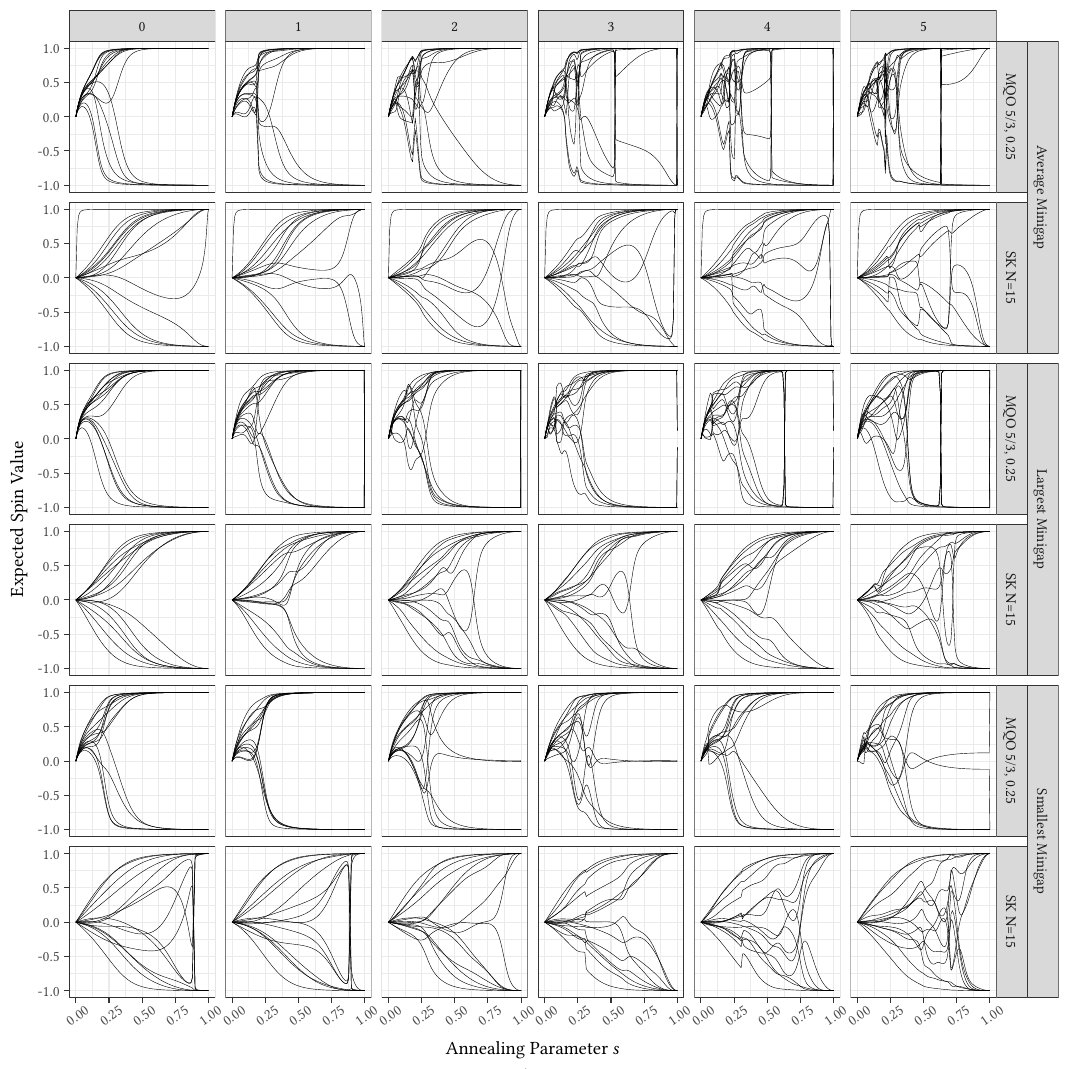}\vspace*{-0.5em}
    \caption{
    Spin-resolution view of temporal annealing dynamics for various instances of Multi-Query Optimisation (MQO) and the Sherrington-Kirkpatrick (SK) Hamiltonian.  
    Each line corresponds to the expected value \(\langle\sigma_{i}^{z}(s)\rangle\), the instantaneous \(z\)-magnetization of one spin. Values near 0 indicate \enquote{undecided qubits}, (in physical terminology, this corresponds to \emph{frustration}), while values near \(\pm\) 1 indicate qubits that have effectively converged to a classical assignment observed in the measurement process at the end of the annealing run.
    Smooth trajectories correspond to gradual bit fixation; abrupt collective changes signal re-organisation between competing low-energy configurations, typically near small spectral gaps. Comparing panels shows that smaller minimum gaps correlate with later, sharper, and more non-monotone spin commitments.}\label{fig:spin-dynamics}
\end{figure*}

\begin{figure*}[htbp]
    \includegraphics{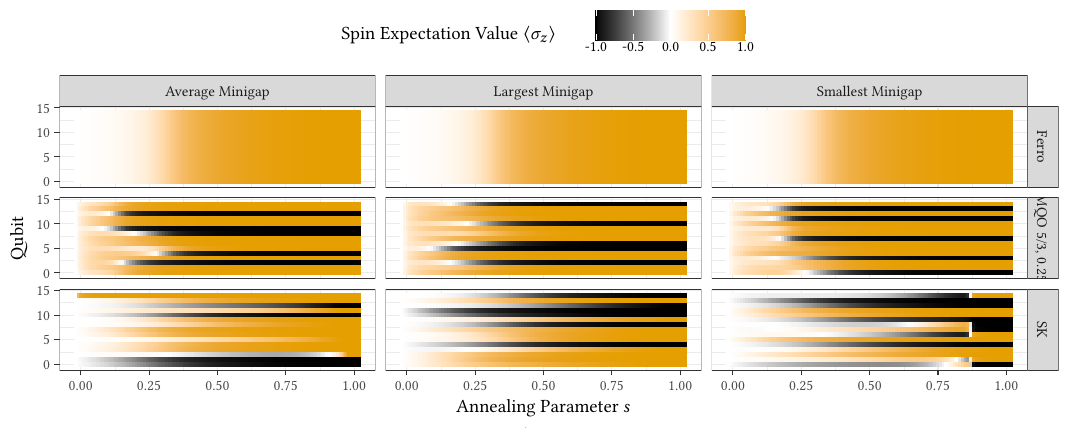}\vspace*{-0.5em}
    \caption{Development of expected values of spins settings. Each row corresponds to the expected value \(\langle\sigma_{i}^{z}(s)\rangle\).
    Note that since the ferromagnetic Hamiltonian does not admit different instances, the data are identical across all three occurrences, which we keep as a comparison reference for an easy to solve problem on a quantum annealer.}\label{fig:spin-expectation}
\end{figure*}

The energy spectrum has important practical consequences: Knowing regions with small and large gaps (especially since different problem instances often show similar structural properties~\cite{Krueger:2025}) allows us to (a) construct 
\emph{concrete} annealing schedules \(s(t)\) that proceed faster for large and slower for asmall gaps. This minimises unwanted transitions (\enquote{leaks}) into 
higher energy levels; and (b) predict \emph{ideal} performance on future noiseless machines as reference against existing approaches. It is (c) also the basis for HW-SW-co-design~\cite{Safi:2023} approaches (\eg, adaptive error correction), or (d) semi-classical approaches based on effective physical properties.

While the spectrum is hard to compute itself, it is know to behave similarly 
across instances of a given problem, and knowledge for one instance
can inform the choice of settings for others.

\subsection{Annealing Dynamics}
\autoref{fig:characteristic-dynamics} analyses the annealing dynamics of some
of our representative Hamiltonians through the quantity (usually referred to
as adiabatic condition ratio or adiabatic parameter~\cite{Roland2002, Farhi2000, Jansen2007})
\begin{equation}
R(s)=\frac{|\bra{E_{1}}\partial_{s}\hat{H}\ket{E_{0}}|}{\Delta_{10}(s)^2},
\end{equation}
that directly governs the local rate of diabatic transitions during quantum annealing (the plot also separately shows the constituents of the fraction;   
note that since the the smallest minigap for SK 
leads to a near-singular value of \(R\) that makes
it hard to reconcile the visualisation with the two other cases, we omit this instance in the plot). While the minimum spectral gap \(\min_{s} \Delta_{10}(s)\) is frequently used as simple proxy for problem hardness, \(R(s)\) provides a more informative diagnostic. For the ferromagnetic Hamiltonian, \(R(s)\) exhibits a single, narrow peak roughly centred near the quantum critical point, with a modest tail extending toward larger \(s\)) (note that since we only consider a
single parametrisation for a ferromagnet, the same profile is shown in both comparisons, and contains identical information). This reflects a  second-order quantum phase transition accompanied by a polynomially closing gap; recall that this gap behaviour
can also be analytically ascertained. The localisation of the peak implies that diabatic transitions are largely confined to a small region of the annealing schedule. This indicates that schedule optimisation on practical devices~--~such as locally slowing the evolution near the critical point~--~is effective, rendering these problem instances comparatively easy for quantum annealing. In contrast, the Sherrington–Kirkpatrick (SK) spin glass displays qualitatively richer behavior. \(R(s)\) profiles remain broad and show significant weight both early and late in the anneal. This structure arises from multiple avoided crossings generated by a rugged energy landscape with many competing low-energy configurations. Diabatic transitions are not concentrated at a single bottleneck, but distributed across a wide interval of \(s\), making the annealing process more challenging. In particular, the late-anneal features reflect residual avoided crossings among the nearly degenerate classical configurations that proliferate in the spin-glass phase.

The MQO problems display behaviour between the two extremes of a known easy and a known hard problem: \(R(s)\) is generally broader and less symmetric than the ferromagnetic case, and also shows more similarity between easy and typical instances (a more
detailed analysis of the consequences for data management is left to follow-up work). Nonetheless, the figure underlines that annealing hardness cannot be reduced to the minimum gap alone, and emerges from the interplay between spectral gaps and associated transition matrix elements distributed across the entire annealing path: this information is not available from straightforward empirical analysis of experimental runs on prototype devices.

\begin{figure*}[htbp]
    \includegraphics{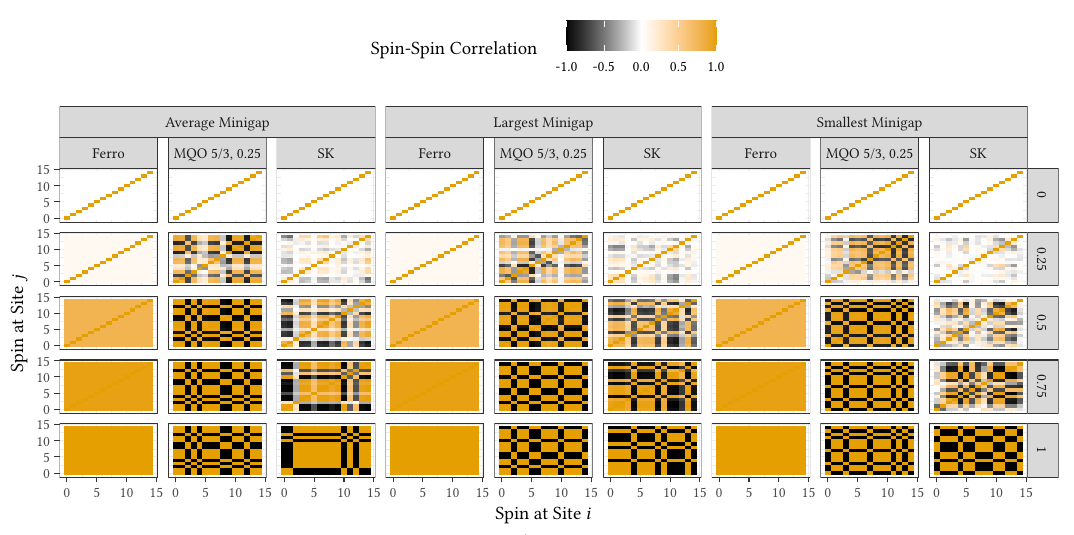}\vspace*{-0.5em}
    \caption{Spin-Spin Correlations. Each field 
    in the grid represents the expected value \(\langle\sigma_{i}^{z}\sigma_{j}^{z}\rangle\).
    While detail are discussed in the text, observe the particularly salient plateau with \enquote{undecided} (\ie, highly fluctuating) spins that appears in
    white colour, corresponding to expectation value 0, for the \emph{smallest} minimum gap instance of MQO; the corresponding phenomenon is already visible for the SK problem with an \emph{average} minimum gap. The ferromagnet discussion in~\autoref{fig:spin-expectation} applies.}\label{fig:spin-spin-correlation}
\end{figure*}

\subsection{Spin Dynamics and Correlations}
\autoref{fig:spin-dynamics} shows temporal dynamics of spins how (and thus: optimisation variables) evolve into their final measured states. As they can be in a superposition state, we use \emph{expected} values \(\langle \sigma_{i}^{z}\rangle\) over normalised annealing time. MQO instances exhibit early and largely monotonic polarisation (\ie, variables assume distinct values) of individual spins, indicating that the instantaneous ground state rapidly approaches a product state in the computational basis that gives unique measurement results. In contrast, SK instances show extended regions where 
\(\langle \sigma_{i}^{z}\rangle \approx 0\), followed by abrupt and collective transitions. This reflects the highly frustrated, dense interaction structure of SK, where low-energy configurations differ globally and spins cannot be fixed independently. MQO behaves like a weakly coupled problem amenable to incremental decision-making, whereas SK requires coordinated global arrangements, consistent with its greater computational hardness (as usual, a detailed discussion of the consequences is left to follow-up work, as the aim of this paper is to introduce methods and computational techniques for the required analysis).

Note that we allow for deliberately tailoring the Sherrington-Kirkpatrick Hamiltonian such that one spin is effectively pinned to the up state by setting the field strength respectively coefficient appropriately.
This removes exact pairwise degeneracy between globally flipped configurations. Without fixing a spin, observables odd under spin flip satisfy
\(\langle \sigma_{i}^{z}\rangle = 0\) in any symmetry-preserving ground state (or a thermal mixture), even when the system is effectively choosing between two opposite classical solutions. Fixing one spin selects a symmetry sector, so the remaining
\(\langle \sigma_{i}^{z}\rangle\) can become non-zero and reveal the structure of the chosen solution branch. Also, it can improve numerical stability and interpretability~\cite{Mezard1986}, as near-degenerate symmetric states can mix arbitrarily in finite-precision
simulations, causing sign flips or basis-dependent behaviour from run to run. Importantly, it also adds a plausibility check to ascertain the correctness of our simulation: One spin in the average case instance simulation for SK (we apply the tailoring for this instance) immediately converges to polarisation state +1, and remains in this state throughout the complete temporal evolution because of this pinning. This is visible in the top left trajectory.

Similar insights can be obtained using a different
 representation shown in \autoref{fig:spin-expectation} and 
\autoref{fig:spin-spin-correlation}: These might be more appealing to a computer science audience, as they resemble correlation and mean-value plots that are commonly used in data science. 
It is interesting to observe that the correlation matrices develop block/checkerboard patterns
that sharpen with increasing \(s\); the visualisation again hints that MQO evolution is 
more structured than SK, but less trivial than for the ferromagnet. The blocks likely correspond to logical clusters or constraints in the 
optimisation problem, whereas correlations may
reflect shared variables or mutual exclusion constraints.

\section{Related Work}\label{sec:related}
Research at the intersection of quantum computing and data management uses quantum or quantum-inspired hardware to accelerate classical database tasks such as query optimisation, transaction scheduling, schema matching, or index tuning.
Cost-based query optimisation has become one of the main testbeds for quantum methods because many of its core subproblems are NP-hard and admit compact combinatorial encodings~\cite{Trummer2016,Fankhauser2021,Fankhauser2023}.

Join ordering is arguably the best-developed subarea within quantum query optimisation;
following initial QUBO formulations~\cite{Schoenberger2023ready}, quantum-inspired digital annealing~\cite{Schoenberger2023} and broader approaches from left-deep to bushy plans~\cite{Nayak2023,Schonberger2023General,Trummer2025} have been considered, as well as quantum machine learning based methods~\cite{Winker2023,Franz2024,Liu2025}.
Additionally, transaction scheduling~\cite{Bittner2020,Bittner2020Hardware,Groppe2021}, schema matching~\cite{Fritsch2023}, index tuning and advising~\cite{Gruenwald2023,Kesarwani2024}, and
multi-query optimisation~\cite{Trummer2016,Schoenberger2025}.
have been considered; see also the survey in Ref.~\cite{Calikyilmaz2023}.

While latest experimental results show clear signs of beyond-classical power of quantum annealing~\cite{King2025}, a comprehensive theory of the exact relationship between quantum phase transitions~--~known to be key in quantum annealing~\cite{Amin2009, Young2010, Amin2009b}~--~and parameter regimes/conditions under which quantum annealing fails if not yet know~\cite{Werner2023}. Performance of quantum annealing itself has been subject to considerable scrutiny~\cite{Mehta2021,Gabor2019,Krueger2020,Sax2020,Mehta2025}, and efforts to determine the
salient properties like minimum spectral gap of problems without solving the problem itself have been made~\cite{Bode2024}.

\section{Conclusion}\label{sec:conclusion}
\looseness-1 We presented a computational framework for analysing combinatorial optimisation problems addressed via quantum annealing. By emphasising principled methods for evaluation, visualisation, and interpretation, we aim to provide a foundation for more rigorous and meaningful analyses of quantum approaches within the DBMS community. We anticipate that such a perspective will support more reliable assessment of potential advantages and inform future developments at the intersection of data management and quantum computing.

\begin{small}
\textbf{Acknowledgements} This work was partly supported by the German Research Foundation, grant MA 9739/1-1, and by the High-Tech Agenda of the Free State of Bavaria. We also acknowledge partial support by the European Union (Project Reference 101083427) and the European Funds for Regional Development (EFRE) (Project Reference 20-3092.10-THD-105). 
\end{small}


\FloatBarrier
\bibliographystyle{ACM-Reference-Format}
\bibliography{references}
\end{document}